\documentclass{ws-procs975x65}

\newcommand{\beq}{\begin{equation}}
\newcommand{\eeq}{\end{equation}}

\newcommand{\be}{\begin{equation}}
\newcommand{\ee}{\end{equation}}

\newcommand{\bea}{\begin{eqnarray}}
\newcommand{\eea}{\end{eqnarray}}
\newcommand{\non}{\nonumber}

\begin{document}

\title{Quasi Scale-Invariant Inflationary Attractors}
\author{Massimiliano Rinaldi$^*$ }

\address{Department of Physics, University of Trento,\\
via Sommarive 14, 38123 Trento, Italy\\
and\\
INFN - TIFPA\\via Sommarive 14, 38123 Trento, Italy\\
$^*$E-mail: massimiliano.rinaldi@unitn.it\\
}

\begin{abstract}
We show that pure quadratic gravity with quantum loop corrections yields a viable inflationary scenario. We also show that a large family of models in the Jordan frame, with softly-broken scale invariance, corresponds to the same theory with linear inflaton potential in the Einstein frame. It follows that all these quasi scale-invariant models have the same  relation between the tensor-to-scalar ratio and the scalar spectral index, which is also consistent with the current data.  Thus, they form a family of attractors, which is sharply distinct from the recently discovered $\alpha$-attractors of Kallosh, Linde et al.
\end{abstract}
\vspace{1cm}

\bodymatter


\noindent  The status of inflationary cosmology has entered, for the first time, a phase where observational data are so stringent that many known theoretical models  can be safely excluded \cite{P15,B14,B15}. Therefore,  since inflation is most likely a quantum gravitational phenomenum, new cosmological observations have the potential to shed some light on the interplay between gravity and quantum field theory in the early Universe.

In this work we consider the family of  modified gravity theories characterized by the Lagrangian
\bea\label{lagra}
L=\sqrt{g} f(R)\,,
\eea
where $f(R)$ is a function of the Ricci scalar $R$  \cite{defelice}. In the framework of inflationary theory, one of the most popular models of this sort was proposed by Starobinski \cite{staro}. In his model, the Lagrangian explicitly reads $f(R)=R+R^{2}/(6M^{2})$, where $M$ is a mass scale,  and it yields predictions on the scalar spectral index $n_{s}$ and the scalar-to-tensor ratio $r$ that are compatible with the latest data.

Our analysis begins with the fact that, while the experimental value of $n_{s}=0.968\pm 0.006$ is known with great accuracy,  the one of $r$  ranges in the interval $0\leq r \lesssim 0.1$  \cite{P15}. This uncertainty on $r$ yields a great degeneracy of models. For example,  in single-field inflation, any potential of the form $V\sim \phi^{p}$ with  $0<p\leq 2$ can explain these data with very good accuracy. A similar situation exists in $f(R)$ gravity since the measured values of the spectral indices can be explained with an effective Lagrangian of the form 
\bea\label{R2delta}
f(R)=R^{2-\delta}\,,
\eea
where  $0<\delta\ll1$ \cite{tn1}. Since $f(R)=R^{2}$ corresponds to an exactly scale-invariant theory, this result leads to the conclusion that, during inflation, the underlying gravitational theory can be described as a ``softly broken'' scale-invariant model. The pure quadratic gravity Lagrangian $L=\sqrt{g}R^{2}$ has very special properties. At the classical level, its equations of motion in vacuum admit an isotropic and homogeneous solution that smoothly interpolates between a radiation-dominated Universe and a de Sitter space with arbitrary cosmological constant. It also admits spherically symmetric black hole solutions with topological horizons and scale-invariant thermodynamical laws \cite{Kehagias,tn2,tn5}. At the quantum level, quadratic gravity is loop renormalizable and ghost-free \cite{koun}, therefore this model is particularly attractive as a playground for inflationary cosmology.

A possible physical interpretation of the result \eqref{R2delta} can be found in the realm of semiclassical theory. Loop corrections to quadratic gravity with a de Sitter background can be computed via functional methods, and the result has the form \cite{loopcorr}
\bea\label{fullaction}
f_{\rm eff}(R)=R^{2}\left[1-\gamma\ln\left(R^{2}\over \mu^{2}\right)\right]\,.
\eea
Since the quantum correction is supposed to be small, one might be tempted to identify the $\delta$-correction in \eqref{R2delta} with the loop logarithmic contribution above. However, it can be shown that this identification yields wrong predictions, in particular it leads to $n_{s}>1$. Therefore, it seems that one loop-corrected quadratic gravity alone cannot describe inflation. One possibility then is that some matter field is required, in the form of additional scale-invariant operators, which break the symmetry via quantum corrections \cite{strumia}. 

In alternative, and in analogy with QCD, it might be that inflation is a non-perturbative phenomenon and this is the reason why the one-loop corrected Lagrangian \eqref{fullaction} gives wrong results. Along these lines, we have shown that a phenomenologically viable inflationary model is given by the Lagrangian \cite{tn3}
\bea\label{resum}
f_{\rm eff}(R)={R^{2}\over\left[1+\gamma\ln\left(R^{2}\over \mu^{2}\right)\right]}\,,
\eea
which, at the leading order in slow-roll parameters, give the relation
\bea\label{univ}
r={8\over 3}(1-n_{s}).
\eea
Note that this result is independent of $\gamma$, the only free parameter of the theory. In addition, for $n_{s}= 0.968$, this relation yields $r=0.085$, which is compatible with the latest data.

The model \eqref{resum} might seem very \emph{ad hoc}, but, in principle, it can be seen as a resummation of loop-corrections, with a Landau-like pole at $R=\mu\exp [1/(2\gamma)]$, in the spirit of the analogy with QCD mentioned above. But apart from this possible analogy, we have shown that \eqref{resum} is equivalent, in the slow-roll approximation (where we neglect the kinetic term of the scalar field) and on-shell, to the induced gravity model with the Coleman-Weimberg potential of the form \cite{gianni}
\bea
L_J=\sqrt{-g}\left[\xi \psi^{2}\left(1+{\varepsilon\over 2}\ln\left(\psi^{2}\over \mu^{2}\right)\right) R - \frac{1}{2}(\partial \psi)^2- \lambda \psi^{4}\left(1+{3\varepsilon\over 4}\ln\left(\psi^{2}\over \mu^{2}\right)\right)   \right]\,,
\label{gianni}  
\eea
where $|\varepsilon|\ll 1$.

This correspondence is not a mere coincidence. To show this, we  consider the model \cite{tn4}
\bea
L=\sqrt{-g}\left( \frac{M^2}{2} R - \frac{A_p}{2\phi^p}(\partial \phi)^2- V(\phi)\right)\,.    
\label{kl}
\eea
where the kinetic term is non-canonical and has a pole of order $p$. This parametrization is inspired by the so-called $\alpha$-attractors studied by Kallosh, Linde et al. \cite{linde,roest}. In the latter case, the potential $V(\phi)$ is taken to be smooth at the pole of the kinetic term. In our case, we considered instead potentials that are not analytic at the pole and have the prototypical form
\bea\label{proto}
V=V_{0}\left(\phi\over \phi_{0}\right)^{{(2-p)/2}}\left[1+\beta\ln\left( {\phi\over m}\right)\right]\,,\quad p\geq 2\,.
\eea
By calculating the slow-roll parameters for these models, we have found that the spectral index is related to the tensor-to-scalar ratio by the equation
\bea
r&=& {8\over 3}(1-n_{s})-{32(1-n_{s})\beta\over 9(p-2)}+{\cal O}(\beta^{2})\,,\quad p\neq 2\\\non
r&=& {8\over 3}(1-n_{s})\,,\quad p=2\,.
\eea
Thus, for $p=2$, we recover eq.\ \eqref{univ}. This is easily explained by noting that, by a suitable conformal transformation to the Jordan frame, we can map the Lagrangian \eqref{kl} into the form \eqref{gianni} for $p=2+\varepsilon$ and by expanding around $\varepsilon=0$. 

Therefore, we have established that all models with a Lagrangian given by \eqref{kl} and a potential of the form \eqref{proto} yield, at the leading term of the slow-roll parameters, the universal relation \eqref{univ}. By means of a conformal transformation to the Jordan frame, we find that all these model are in fact equivalent, on shell, to the quasi-scale invariant quadratic model \eqref{resum}. 

In conclusion, the so-called concave-convex divide in the $(n_{s}, r)$ plane, which is usually identified with the simple linear inflaton potential, turns out to correspond to a whole class of scale-invariant models implemented by loop corrections. In this sense, the straight trajectory \eqref{univ} is seen as an \emph{attractor}, which, however, is distinct form the $\alpha$-attractors of Kallosh and Linde, characterized by a different relation between $r$ and $n_{s}$. 

The picture that emerges from our work is that of a Universe that originates from the breaking of scale invariance by means of quantum corrections. In principle, the Universe might start in a classical state of radiation domination with vanishing scalar curvature $R$, which is a generic solution for the equations of motion derived from the Lagrangian $L=\sqrt{g} R^{2}$. As we have mentioned above, the Universe evolves naturally from this state towards a de Sitter phase with arbitrary cosmological constant (which is an attractor solution), until quantum loop corrections set in and drive the Universe towards an inflationary phase. The subsequent reheating mechanism is not known yet but it probably relies on the presence in the Lagrangian of other scale invariant operators,  together with their quantum corrections.


\end{document}